\documentclass{svproc}
\usepackage{url}

\usepackage[usenames,dvipsnames]{color}
\usepackage{amsmath,amssymb,mathtools,xspace}
\usepackage{multirow}
\usepackage{graphicx}
\usepackage{adjustbox}
\usepackage{booktabs}
\usepackage{appendix}
\usepackage{slashed}
\usepackage{xspace}
\usepackage{braket}
\usepackage{tabularx}
\usepackage{subcaption}
\usepackage[normalem]{ulem} 
\usepackage{siunitx}
\usepackage[ISO]{diffcoeff}
\usepackage[utf8]{inputenc}
\usepackage{xspace}
\usepackage{bm}
\definecolor{darkblue}{rgb}{0,0,0.5}
\definecolor{darkgreen}{rgb}{0.0,0.5,0.2}
\definecolor{darkred}{rgb}{0.6,0,0}
\usepackage[pdfstartview=XYZ,
bookmarks=true,
colorlinks=true,
linkcolor=darkred,
urlcolor=blue,
citecolor=darkgreen,
pdftex,
bookmarks=true,
breaklinks=true,
linktocpage=true, 
hyperindex=true
]{hyperref}
\usepackage[capitalise]{cleveref}
\usepackage{orcidlink}

\begin{document}
\mainmatter              
\title{Searching for Light Physics at the LHC}
\titlerunning{Light Physics at LHC}  
%
\author{Patrick Foldenauer\,\orcidlink{0000-0003-4334-4228}}
\authorrunning{Patrick Foldenauer} 
\institute{Instituto de F\'isica Te\'orica UAM-CSIC, Universidad Aut\'onoma de Madrid, Calle de Nicol\'as Cabrera 13–15, Cantoblanco, E-28049 Madrid, Spain\\
\email{patrick.foldenauer@csic.es}} 

\maketitle  

\begin{abstract}
Over the last years, new physics in terms of a novel
weakly-interacting massive particle (WIMP) has come more and more under 
pressure from experimental null results. While the remaining WIMP
parameter space will be probed by next generation dark matter
experiments, models of light new physics have become increasingly popular over
the last decade. In an effort to explore the parameter space of such light 
physics, a myriad of custom designed high-precision/low-energy experiments has 
been proposed. In this note, however, I argue that existing LHC 
multipurpose experiments like ATALS and CMS have a so far unexploited potential 
to probe light physics via appearing displaced recoil jets. In the first 
part, I discuss the sensitivity of this signature to 
(ultra-)light scalar and axionic dark matter, while in the second part I
show its sensitivity to high-energy neutrino scattering.
\end{abstract}

\section{Introduction}
\label{sec:introduction}

\textit{This note is a write-up of a talk given by the author at the  `Eighth Workshop on Theory, Phenomenology and Experiments in Flavour Physics (FPCapri2022)' in Villa Orlandi, Capri Island on June 17th 2022. 
The majority of the material presented here is based on work done in Refs.~\cite{Bauer:2020nld,Foldenauer:2021gkm} in collaboration with Martin Bauer, Felix Kling, Peter Reimitz and Tilman Plehn.}

There is strong experimental evidence for physics beyond the Standard Model (SM). Some of the most compelling hints are for example the observation of neutrino oscillations~\cite{SNO:2001kpb}, requiring massive neutrinos, or the latest Planck measurement of the CMB power spectrum, hinting at an overall dark matter (DM) abundance of $\Omega h^2 \sim 0.26$~\cite{Planck:2018vyg}.
Pinning down the correct particle physics model responsible for giving mass to the neutrinos and explaining the effect of dark matter with the correct relic abundance has been the goal of many particle physicists over the past decades. 
Many dedicated experiments have since been built to test interactions of both neutrinos and dark matter in order to gain a better understanding of their enigmatic nature. In this note, however, we want to focus on what complementary information we might gain from studying their interactions at LHC multipurpose experiments like ATLAS or CMS. To this purpose we will investigate the sensitivity of these experiments to ultralight dark matter candidates via appearing displaced jet signature in~\cref{sec:uldm}. In~\cref{sec:nus} we will then argue that a similar search strategy can also be exploited to observe interactions of the copiously produced neutrinos at the LHC.

\section{Ultralight dark matter}
\label{sec:uldm}

Over the last decades the most popular model for DM has been in terms of a weakly-interacting massive particle (WIMP). This has been exhaustively searched for in collider as well as in  direct and indirect detection experiments. While they still remain viable DM candidates, WIMP models have come more and more under pressure from experimental null results~\cite{Leane:2018kjk} and
the remaining available parameter space will be probed by
the next generation of DM experiments~\cite{Arcadi:2017kky}.
Below the  GeV-scale WIMP-window there are many theoretically well-motivated light dark matter candidates,
which can have relic abundances in agreement with observation from production via freeze-in, particle-anti-particle asymmetry, strong dynamics or non-thermal processes like misalignment~\cite{Knapen:2017xzo}. In this section we want to focus on how to search for such light DM candidates.

\subsection{Searches for ultralight dark matter}
\label{sec:uldm_search}

It has been noted that for a particular mass scale of $m\sim 10^{-22}$ eV for the DM candidate the quantum pressure of the DM bose gas counteracts the gravitational collapse and helps to stabilise galaxy-sized DM halos, leading to a better fit of the small-scale matter power spectrum~\cite{Hu:2000ke}. This has lead to increased interest in so-called ultra-light DM (ULDM) models over the last years. 

\begin{figure}
    \centering
    \includegraphics[width=0.46\textwidth]{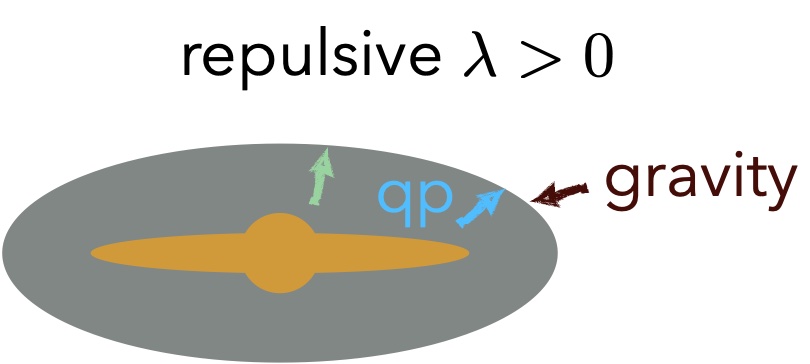}\hfill
    \includegraphics[width=0.46\textwidth]{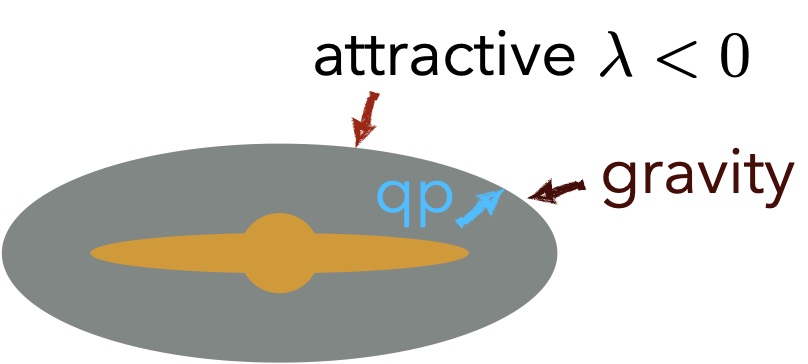}
    \caption{Sketch of DM halo surrounding baryonic matter in galaxy. (Left) Repulsive self-interaction helps to counteract gravity and stabilises halo. (Right) Attractive self-interaction counteracts quantum pressure and destabilises halo.}
    \label{fig:halos}
\end{figure}

However, this picture becomes more subtle as we also considers possible self-interactions of the ULDM particles. 
In case that the ULDM particles are subject to attractive self interactions, $\lambda<0$, these will lead to instabilities of the dark matter halo and ultimately result in its collapse and the formation of boson stars~\cite{Guth:2014hsa}.
On the contrary, if the self-interactions are repulsive ($\lambda >0$) these will help to counteract the gravitational collapse of the halo and hence allow for a relaxed mass range for ULDM of $10^{-22} - 1$ eV~\cite{Ferreira:2020fam}. These two scenarios are sketched in~\cref{fig:halos}.

However, from a theoretical point of view it turns out to be rather difficult to construct well-motivated models for ULDM which naturally lead to repulsive self-interactions~\cite{Fan:2016rda}. Therefore, in a phenomenological approach in this note we will consider two simple extensions of the SM with either a new scalar or pseudo-scalar DM candidate.

\begin{figure}[t]
    \centering
    \includegraphics[width=0.5\textwidth]{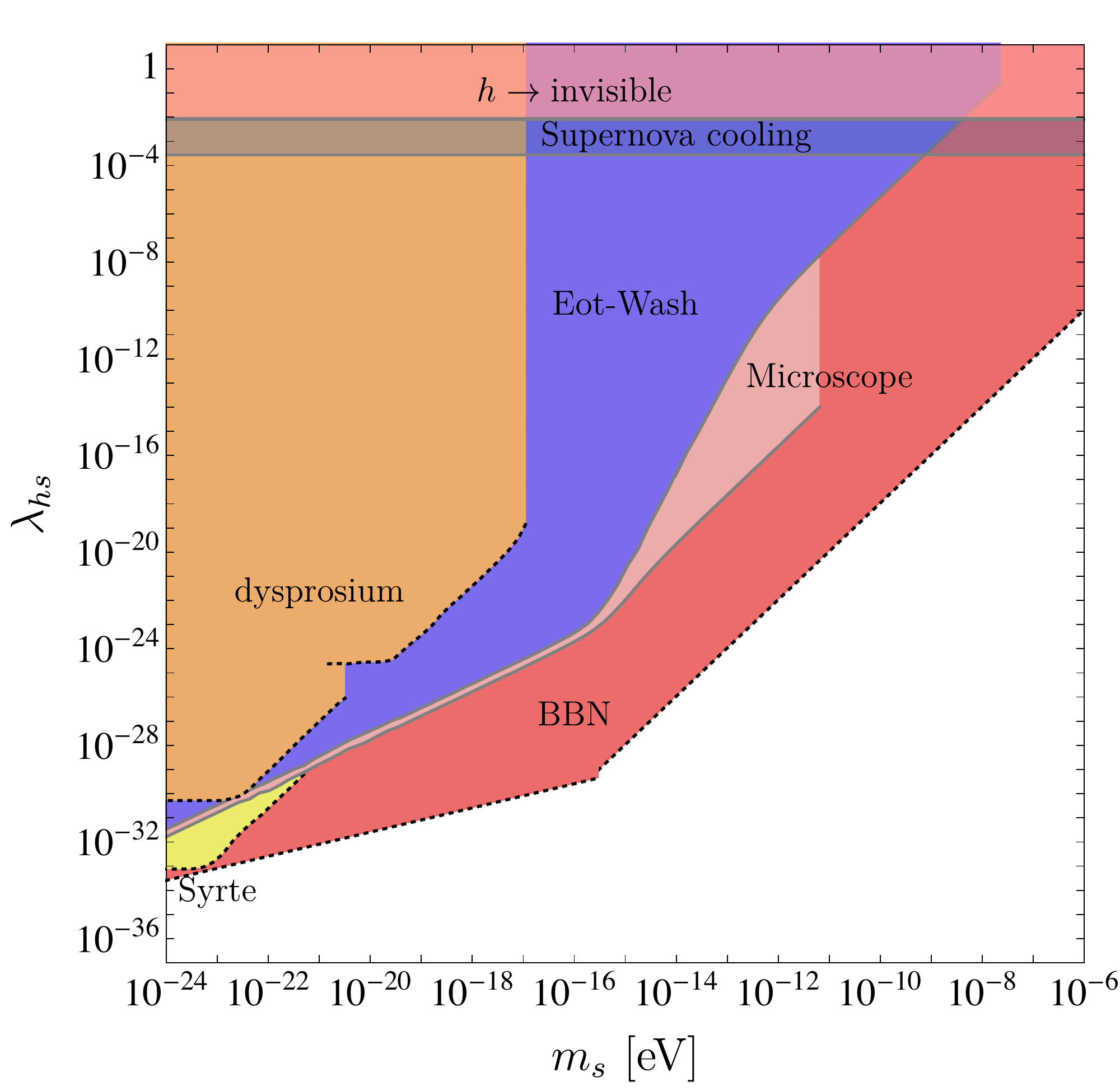}\hfill
    \includegraphics[width=0.49\textwidth]{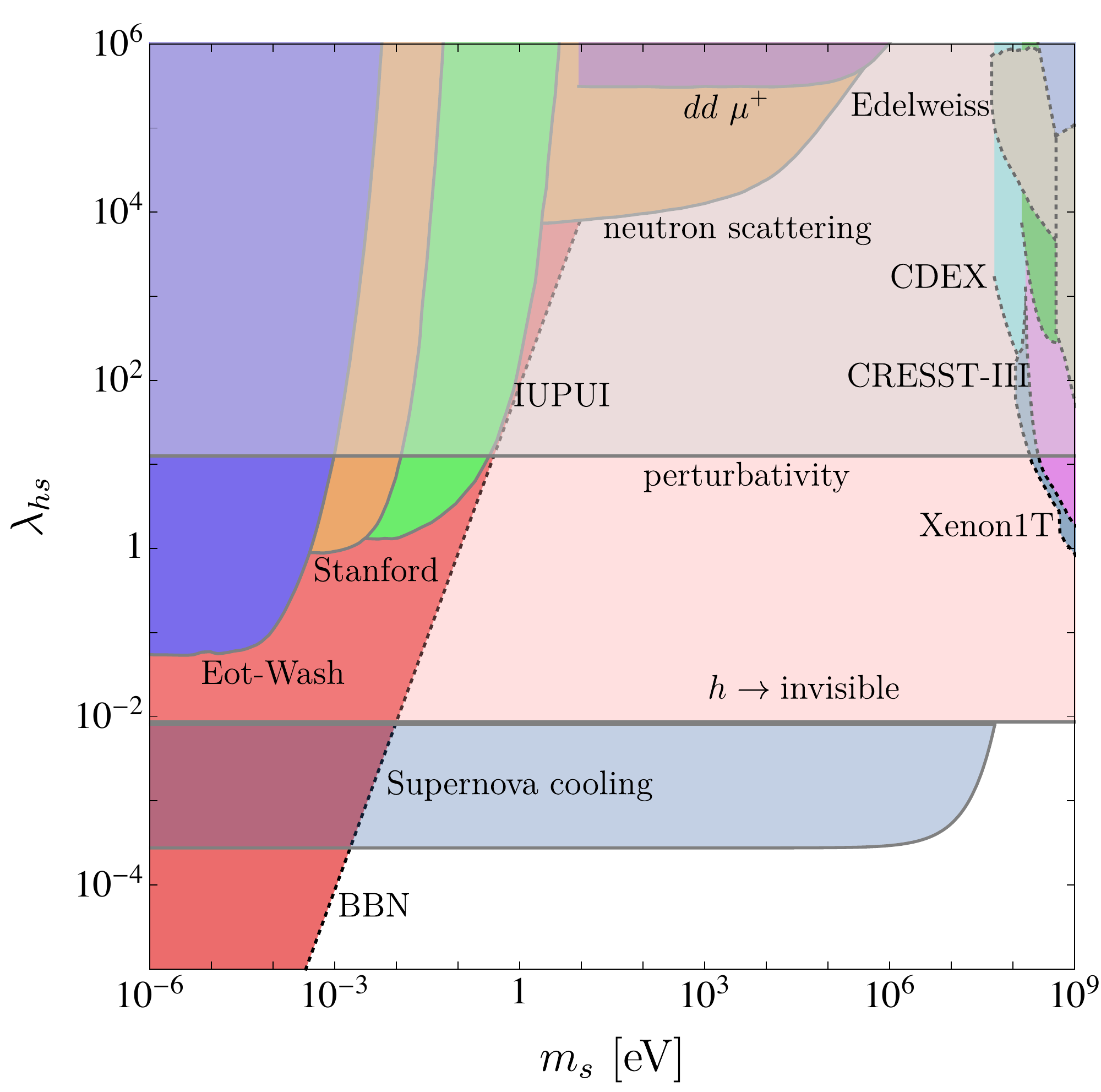}
    \caption{Limits on a new scalar $s$ coupled to the SM via the Higgs portal. Figures taken from~\cite{Bauer:2020nld}.}
    \label{fig:hp_lims}
\end{figure}

The most straight-forward model to consider is that of a real scalar $s$ coupled to the SM via the Higgs portal
\begin{align}
\mathcal{L} \supset -\frac{1}{2}\lambda_{h s}\, s^2\,H^\dagger H\,.
\label{eq:higgsportal}
\end{align}
At low energies, where we can integrate out the Higgs, this interaction leads to an effective DM-nucleon coupling of the form
\begin{align}
\mathcal{L}\supset c_{sNN}\, s^2 \bar N N \,,
\label{eq:nucluoncoupling}
\end{align}
with a coupling coefficient 
\begin{equation}
    c_{sNN} = \lambda_{h s} \frac{m_N}{m_h^2}\frac{2n_H}{3(11-\frac{2}{3}n_L)} \,.
\end{equation}
This type of coupling leads to a potential between nuclei via the exchange of pairs of very light scalars $s$. Hence, it induces a fifth force that can be tested in precision measurements of low-energy observables in e.g.~neutron scattering, molecular spectroscopy or Eot-Wash experiments~\cite{Brax:2017xho}. Furthermore, since cold light DM has extremely high occupation numbers, it can be treated as a classical, coherently oscillating background field, which allows to separate the quadratic scalar interaction into a constant and a time-dependent part of the form
\begin{align}
s^2&= s_0^2 \cos^2(m_st) \to \frac{s_0^2}{2} \left( 1+\cos(2m_st) \right)\,.
\end{align}
Ultimately, these terms will induce variations of fundamental constants~\cite{Stadnik:2015kia,Hees:2018fpg}, or modifications of the primordial helium yield at the time of BBN~\cite{Bauer:2020nld}. 

For the simple Higgs portal model these constraints are collected for ultralight scalars in the left panel of~\cref{fig:hp_lims}. At higher masses such scalar DM candidates are mostly constrained by supernova energy loss, DM direct detection experiments~\cite{Bauer:2020nld} and the search for Higgs to invisible decays at LHC~\cite{Bernaciak:2014pna}. Limits in the high mass regime are collected in the right panel of~\cref{fig:hp_lims}.

\subsection{Ultralight dark matter at the LHC}

\begin{figure}[t]
    \centering
    \includegraphics[width=0.5\textwidth]{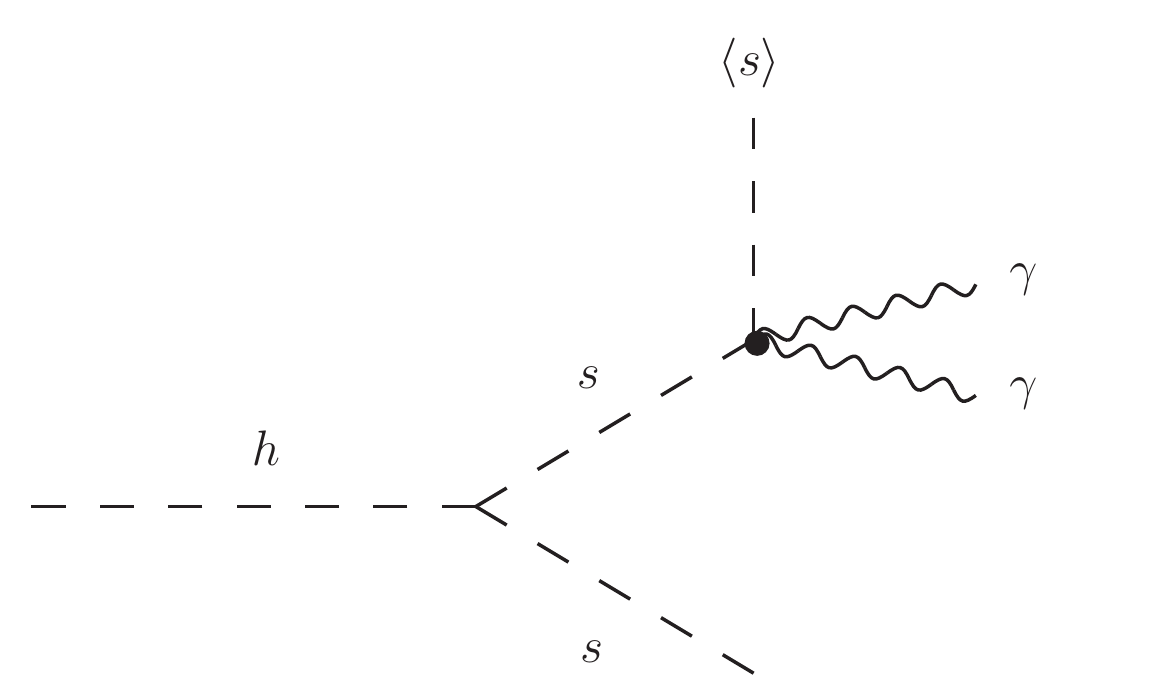}\hfill
    \includegraphics[width=0.5\textwidth]{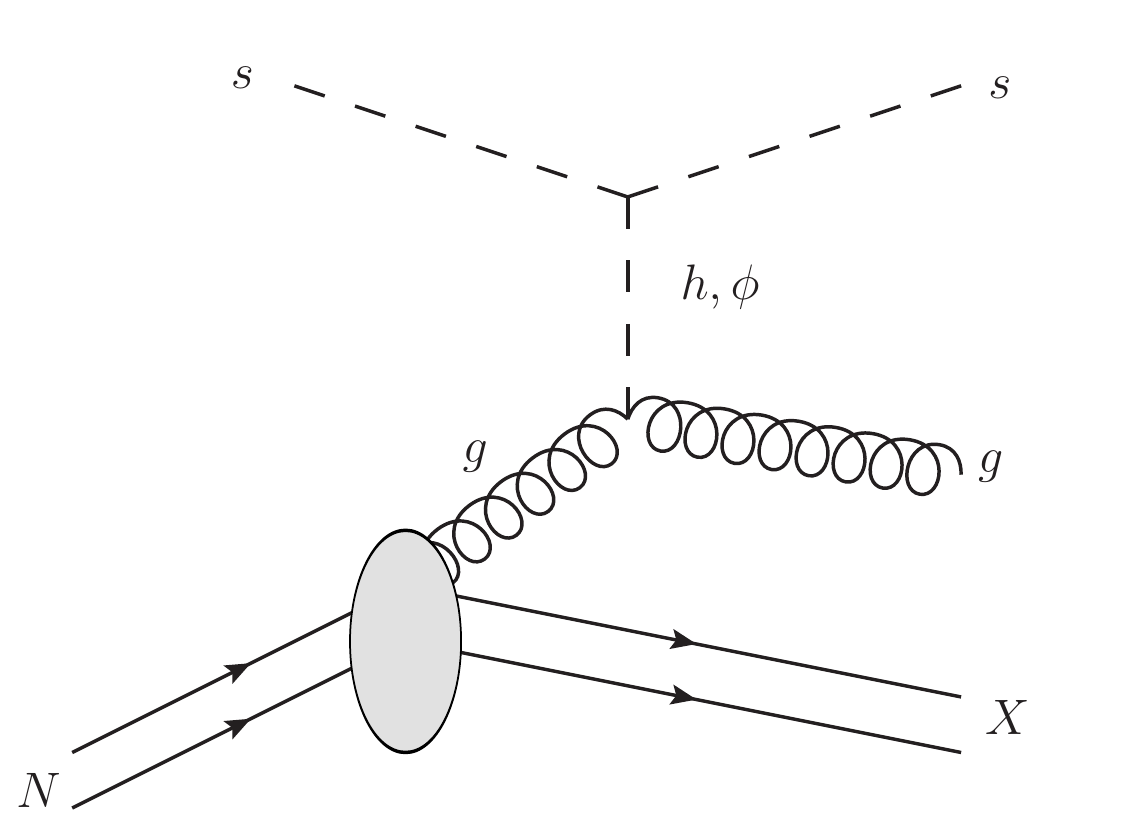}
    \caption{(Left) Higgs decay into two ULDM scalars $s$ with subsequent scattering of the boosted scalar with the DM background. (Right) ULDM-nucleus scattering of a real scalar $s$ with the gluons via the exchange of a Higgs $h$ (or new scalar mediator $\phi$).}
    \label{fig:uldm_lhc}
\end{figure}

In this work we want to evaluate the potential of directly testing interactions of ULDM candidates at the LHC. To this purpose we are studying the sensitivity of LHC multipurpose experiments to two potential detection processes: \textit{(i)} DM-background scattering and  \textit{(ii)} DM-nucleus scattering.

\paragraph*{(i) DM-background scattering}
Staying with the case of the simple scalar Higgs portal model, we can produce the ULDM candidate $s$ at the LHC from Higgs decays. The very boosted DM particle could then undergo scattering with the DM background field in the detector and e.g.~produce a pair of visible photons via its Higgs interactions. This process is illustrated in the left panel of~\cref{fig:uldm_lhc}.

We can compute the mean free path  $\lambda = 1/n_\text{DM} \, \sigma$ for this type of process to get an estimate on which length scales these type of interactions are taking place. Assuming the maximum allowed coupling from Higgs to invisible decays $\lambda_{hs}=8.7\cdot10^{-3}$, 
we arrive at a mean free path for this process of~\cite{Bauer:2020nld}
\begin{align}
\lambda
=  \frac{4 \pi}{\lambda_{hs}^2 g_{h\gamma\gamma}^2} \, \frac{m_h^3}{\rho_\text{DM}} \gtrsim   10^{43}\  \text{m}\,.
\end{align}
This is larger than the size of the observable universe by a factor of $10^{16}$! Hence, this process is entirely unobservable and can be discarded.

\begin{figure}[t]
    \centering
    \includegraphics[width=0.6\textwidth]{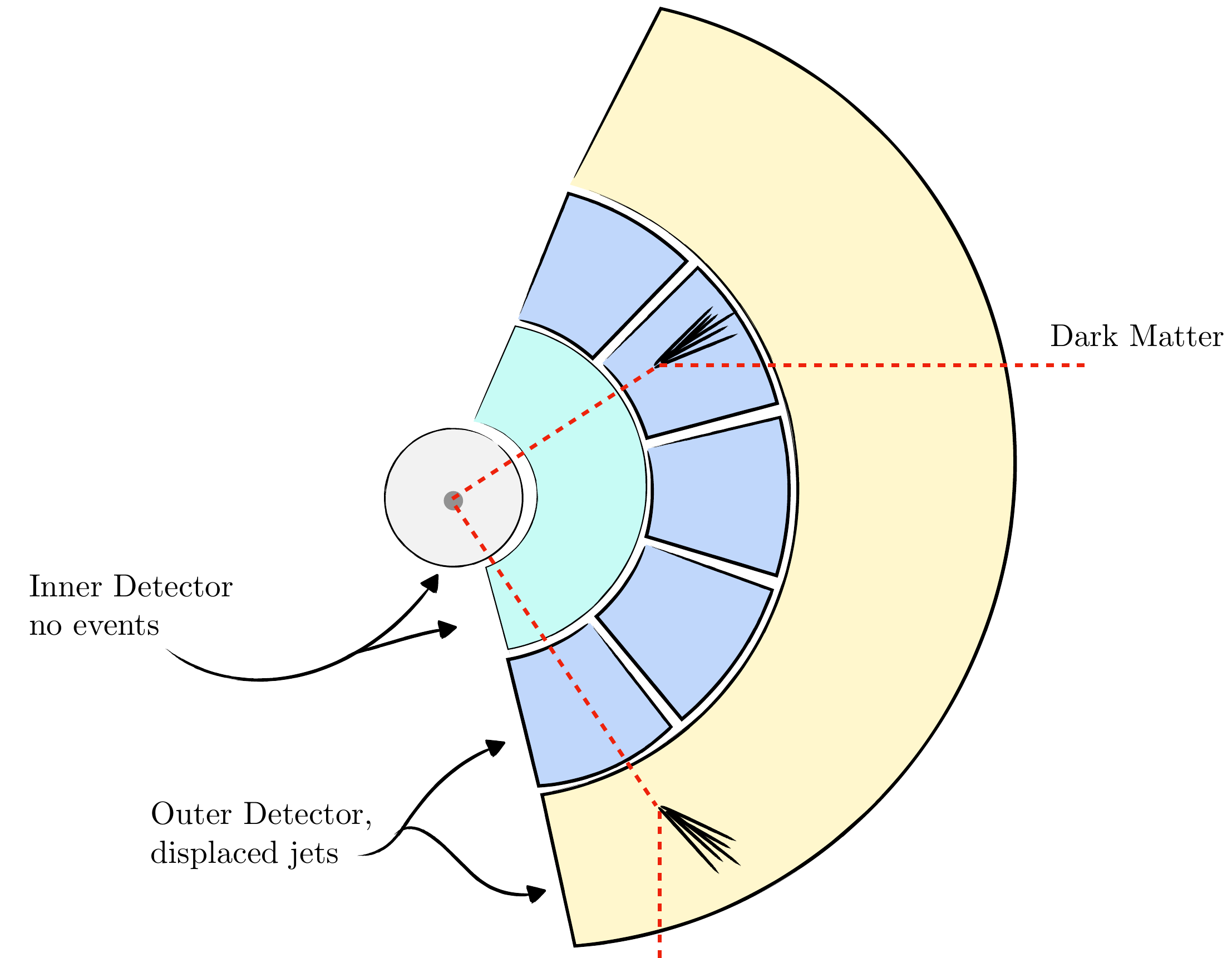}
    \caption{Appearing displaced jet signature in an ATLAS-like detector from scattering of an ULDM particle. Figure taken from~\cite{Bauer:2020nld}.}
    \label{fig:dis}
\end{figure}

\paragraph*{(ii) DM-nucleus scattering}
Alternatively, we can  consider the same production mechanism of the ULDM particles via Higgs decay, but now consider scattering with the detector material itself via the Higgs portal interaction. This process is depicted in the right panel of~\cref{fig:uldm_lhc}.

We can compute the interaction probability of the DM interacting in a detector of length $L_X$ and nuclear density $n_X$ via the mean free path $\lambda_X= 1/n_X \, \sigma_X$,
\begin{align}
P_\text{int}
= 1- e ^{-  L_X/\lambda_X} = 1- e ^{-  L_X\, n_X \, \sigma_X} \; .
\label{eq:int_prob}
\end{align}
Hence, in this process the interaction probability crucially depends on the density of the detector material. In~\cref{fig:dis} we show the typical interaction signature at an ATLAS-like detector. Since the inner parts of the detector are gaseous xenon trackers, the material density $n_{Xe}$ and hence the interaction probability is very low. However, the outer calorimeter parts are mainly made of lead and iron with much higher densities, $n_{Pb} \gg n_{Xe}$. Thus, the DM particle can scatter in these outer, denser detector components and produce a displaced recoil jet~\cite{Bauer:2020nld}. 
The partonic DM-nucleus scattering cross section is typically expressed via the energy loss $\nu$ of the DM particle and the kinematic variables $x = Q^2/(2M\nu)$ and $y = \nu/E_s$,
\begin{align}
\frac{ d^2 \hat\sigma_\text{DIS}}{dx\, dy} = \frac{\lambda_{hs}^2
  g_{hgg}^2}{4 \pi \, \hat s } \; \frac{Q^4 }{(Q^2 + m_h^2)^2} \; ,
\end{align}
where $\hat s = x s = 2 M E_s\, x$ and $Q^2=2 M E_s\, x\, y$. By means of this cross section we can compute the interaction probability in an ATLAS-like detector as
\begin{align}
P_\text{DIS} 
= 1- e ^{-L_\text{E}\, n_\text{Pb}\, \sigma_\text{Pb}}  e ^{-L_\text{H}\, n_\text{Fe}\, \sigma_\text{Fe}}  
\approx7.5 \cdot 10^{-21}  \; .
\label{eq:pdis}
\end{align}
Hence, for the simple Higgs portal interaction this process is again unobservable even at the HL-LHC which aims to collect $\mathcal{O}(10^8)$ Higgses.

\subsubsection{Axion-like particles}

Beyond the minimal scalar Higgs portal model, axions and axion-lke particles (ALPs) are well-motivated candidates  to play the role of ULDM. 
Therefore, as an alternative phenomenological realisation of ULDM we consider an extension of the SM by a shift-symmetric pseudo-scalar (or axion-like) particle $a$. Such a particle can arise as the angular mode of a complex scalar $S$ acquiring a vacuum expectation value $f$,
\begin{equation}
    S = \frac{s+f}{\sqrt{2}}\, e^{ia/f}\,.
\end{equation}
This angular mode $a$ is protected by a shift symmetry of the type 
\begin{equation}
    e^{ia/f} \to e^{i(a+c)/f} = e^{ia/f}e^{ic/f}\,,
\end{equation}
by a constant shift $c$. 

If these particles are coupled to QCD they will acquire a mass term by an explicit breaking of the shift symmetry via non-perturbative effects leading to a potential for the ALP of
\begin{align}
V(a) = \Lambda^4 \left[1-\cos\left(\frac{a}{f}\right) \right] \simeq \frac{\Lambda^4}{2f^2}\, a^2 + \dots \,.
\end{align}
Hence, this potential generates a mass  for the ALP of $m_a = \Lambda^4/2f^2$, where $\Lambda$ is the QCD scale and $f$ the heavy axion scale. Thus, ALPs can quite generically have very suppressed masses, making them good candidates for ULDM.

\begin{figure}[t]
    \centering
    \includegraphics[width=0.5\textwidth]{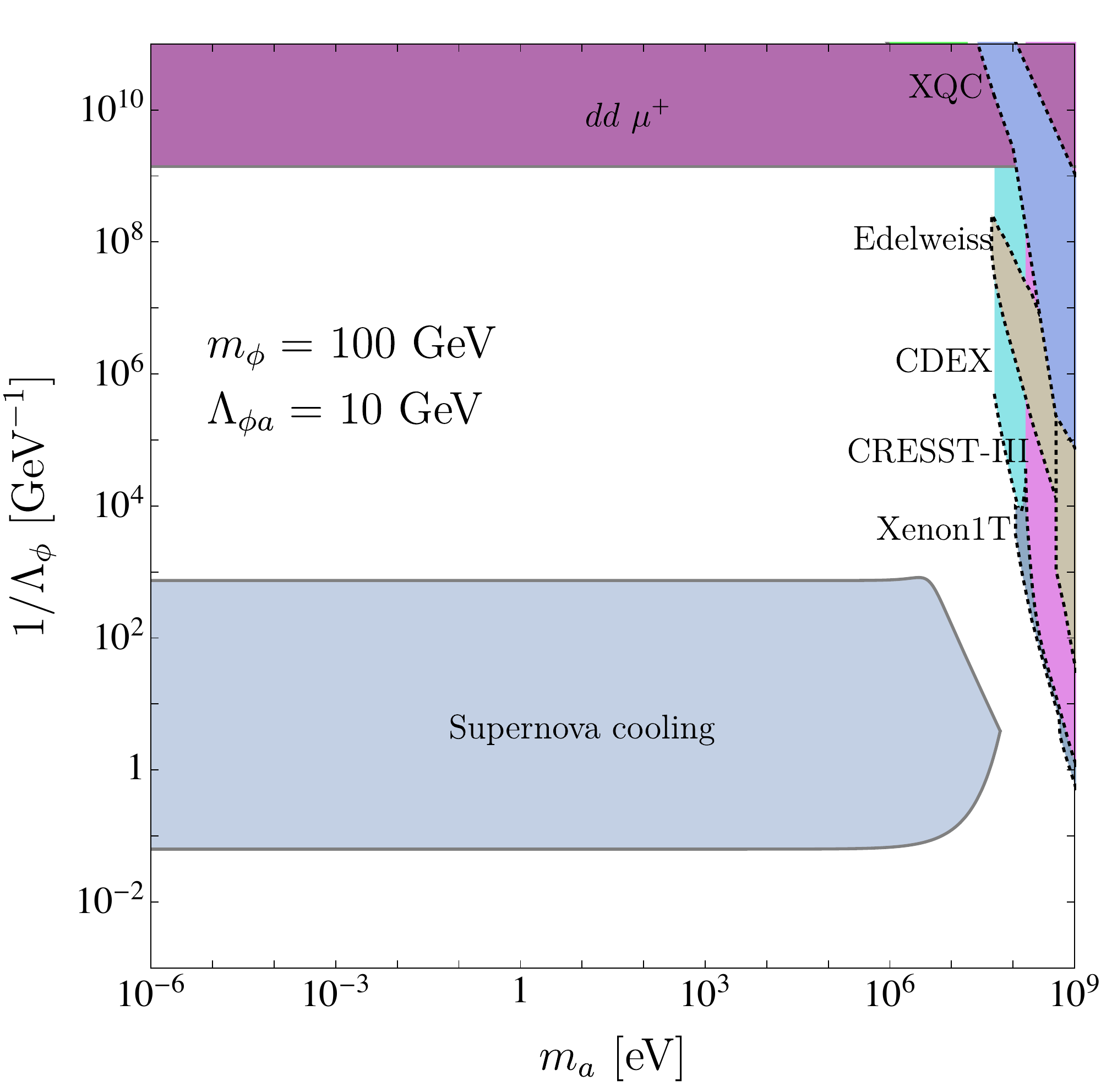}%
    \includegraphics[width=0.5\textwidth]{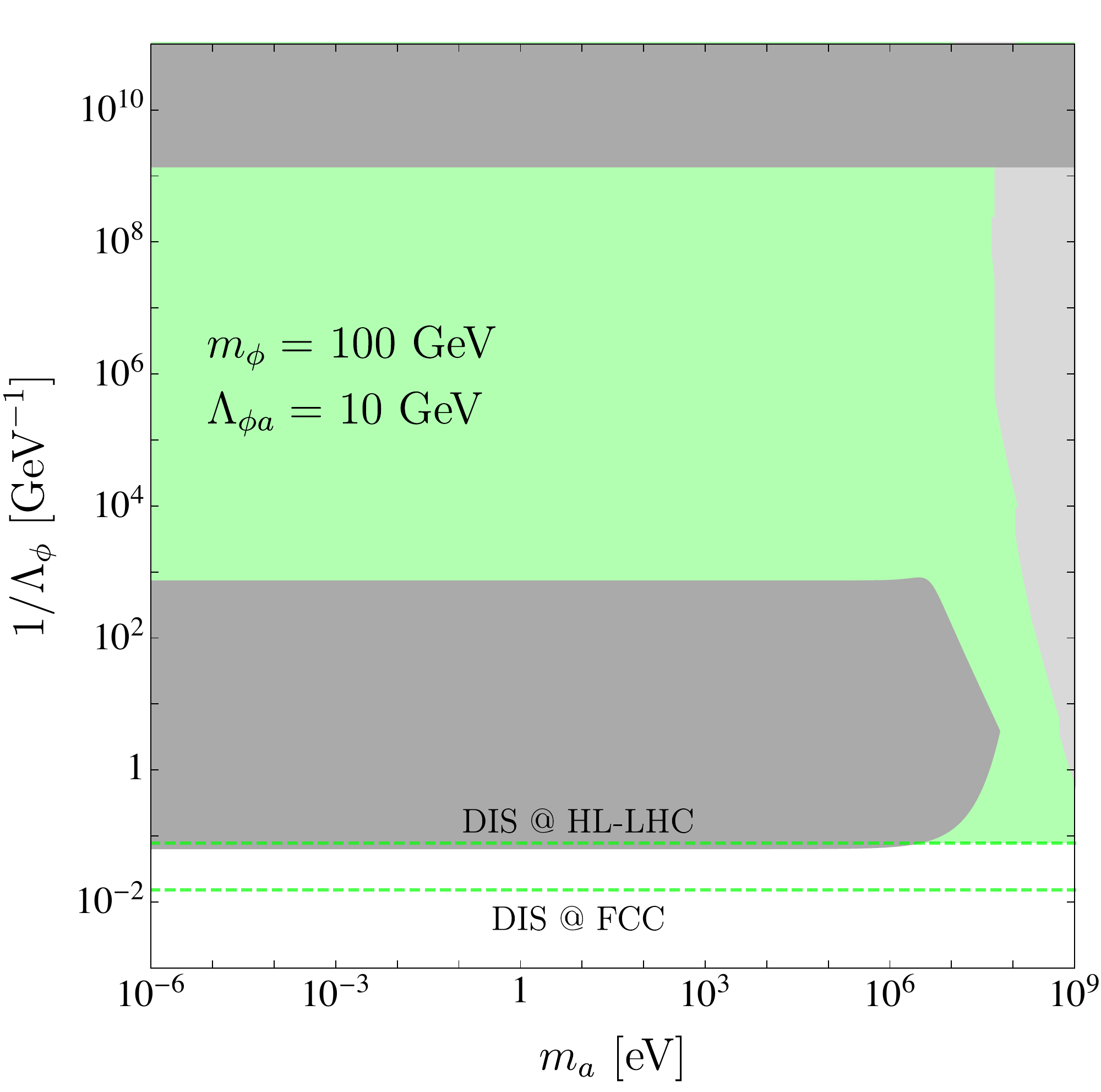}
    \caption{Limits on a new pseudo-scalar $a$ coupled to the SM via a new heavy mediator $\phi$ (left). Appearing displaced jet search sensitivity at HL-LHC and FCC (right). Figures taken from~\cite{Bauer:2020nld}.}
    \label{fig:deriv}
\end{figure}

As a toy model for ultralight ALPs, we consider here a shift-symmetric, pseudo-scalar particle $a$
coupled to the SM via a new heavy scalar mediator $\phi$ via the interaction,
\begin{align}
\mathcal{L} \supset
\frac{1}{2}\partial_\mu a\, \partial^\mu a -\frac{m_a^2}{2}a^2
+ \frac{1}{2}\partial_\mu \phi\, \partial^\mu \phi - \frac{1}{2}m_\phi^2\phi^2
- \frac{\partial_\mu a \,\partial^\mu a}{2\Lambda_{\phi a}}\phi
- \frac{\alpha_s}{\Lambda_\phi} \; \phi \; \mathrm{Tr}\, [G_{\mu\nu}G^{\mu\nu}] \,. 
\label{eq:scalarALPs}
\end{align}
Similar to the simple Higgs portal model, the interactions of~\cref{eq:scalarALPs} lead to an effective low-energy ALP-nucleon coupling,
\begin{align}
\mathcal{L}\supset c_{aNN}\, \partial_\mu a\, \partial^\mu a\ \bar N N \,,
\label{eq:nucluoncoupling2}
\end{align}
with an effective coupling coefficient of~\cite{Bauer:2020nld}
\begin{alignat}{5}
c_{aNN} &=\frac{m_N}{\Lambda_{\phi a}\Lambda_\phi m_\phi^2} \frac{8\pi}{11-\frac{2}{3}n_L}\; .
\label{eq:DerNN}
\end{alignat}
The major difference in the phenomenology of these pseudo-scalar ULDM candidates to scalars is the appearance of the derivative coupling in~\cref{eq:nucluoncoupling2}. This leads to a quadratic scaling of the effective ALP-nucleon coupling  with the momentum transfer $q$ of the interaction. Since all the high-precision searches for ULDM considered in~\cref{sec:uldm_search} are performed at very low energies (compared to the typical weak-scale suppression scales $\Lambda_\phi$, $\Lambda_{\phi a}$) the effective ALP interactions are heavily suppressed compared to the simple real scalar Higgs portal model. This is reflected in the large unconstrained region of parameter space in the left panel of~\cref{fig:deriv}.

These ultralight ALPs can be produced in hadron colliders via the decay of the heavy mediator $\phi$, which is produced via its coupling to gluons in~\cref{eq:nucluoncoupling2}. These ALPs $a$ can scatter in the detector material via the same process we considered in the case of the real scalar $s$. Hence, we can compute the interaction probability of $a$ with the detector via~\cref{eq:int_prob} using the relevant DIS cross section
\begin{align}\label{eq:disDSimp}
\frac{d^2\hat \sigma_\text{DIS}}{dx\, dy} = \frac{\alpha_s^2}{16\pi\, \hat s}\,\frac{Q^4}{\Lambda_{\phi a}^2\Lambda_\phi^2} \, \left( \frac{Q^2+2m_a^2}{Q^2+m_\phi^2}\right)^2\,.
\end{align}
The sensitivities of ATALS at high-luminosity LHC and a future FCC to the resulting displaced jet signature of the
pseudo-scalar ULDM candidate is shown in the right panel of~\cref{fig:deriv}. It can be seen that the LHC search will be able to probe a large part of the currently allowed parameter space and is competitive with bounds from supernova cooling. In the future, FCC will be able to probe couplings which are smaller by even an order of magnitude. The reason why this displaced jet search at high-energy colliders has such a high sensitivity to this model is the derivative interaction of the ALPs $a$. While this leads to a suppression of the interaction rates at low-energy experiments, it results in an enhancement of the signal at colliders.

\section{Neutrinos}
\label{sec:nus}

%
\begin{figure}[t!]
    \centering
    \includegraphics[width=\textwidth]{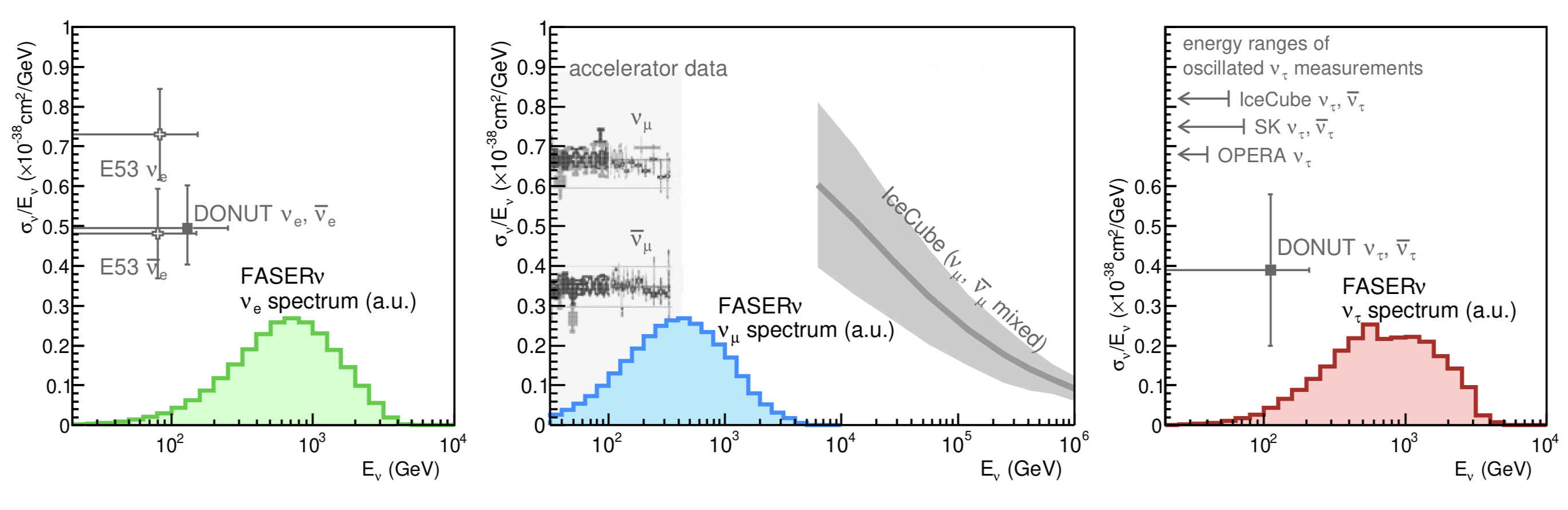}
    \caption{Past neutrino cross section measurements for electron- (left), muon- (centre) and tau-flavoured neutrinos alongside the expected neutrino spectra at
    FASER$\nu$. Figure taken from~\cite{FASER:2020gpr}.}
    \label{fig:faser_nu}
\end{figure}

In the second part of this work we will investigate the sensitivity of a displaced jet search to scattering of SM neutrinos in LHC multipurpose detectors.

While neutrino scattering has been studied in a number of experiments over the last decades, there is still only very limited data available for high-energy interactions. In~\cref{fig:faser_nu} one can see a compilation of neutrino scattering cross section measurements for the different neutrino flavours together with the expected neutrino spectra at the future LHC forward experiment FASER$\nu$. As can be seen we have by far most data for $\nu_\mu$-scattering. However,  at intermediate energies above accelerator experiments and below the high-energy events at IceCube~\cite{IceCube:2017roe} there still is a blind spot to $\nu_\mu$-scattering
at energies between 370 GeV and 6.3 TeV. For $e$- and $\tau$-flavoured neutrinos there is still no data above $\mathcal{O}(100)$ GeV energies at all. 

\begin{figure}[t]
    \centering
    \includegraphics[width=\textwidth]{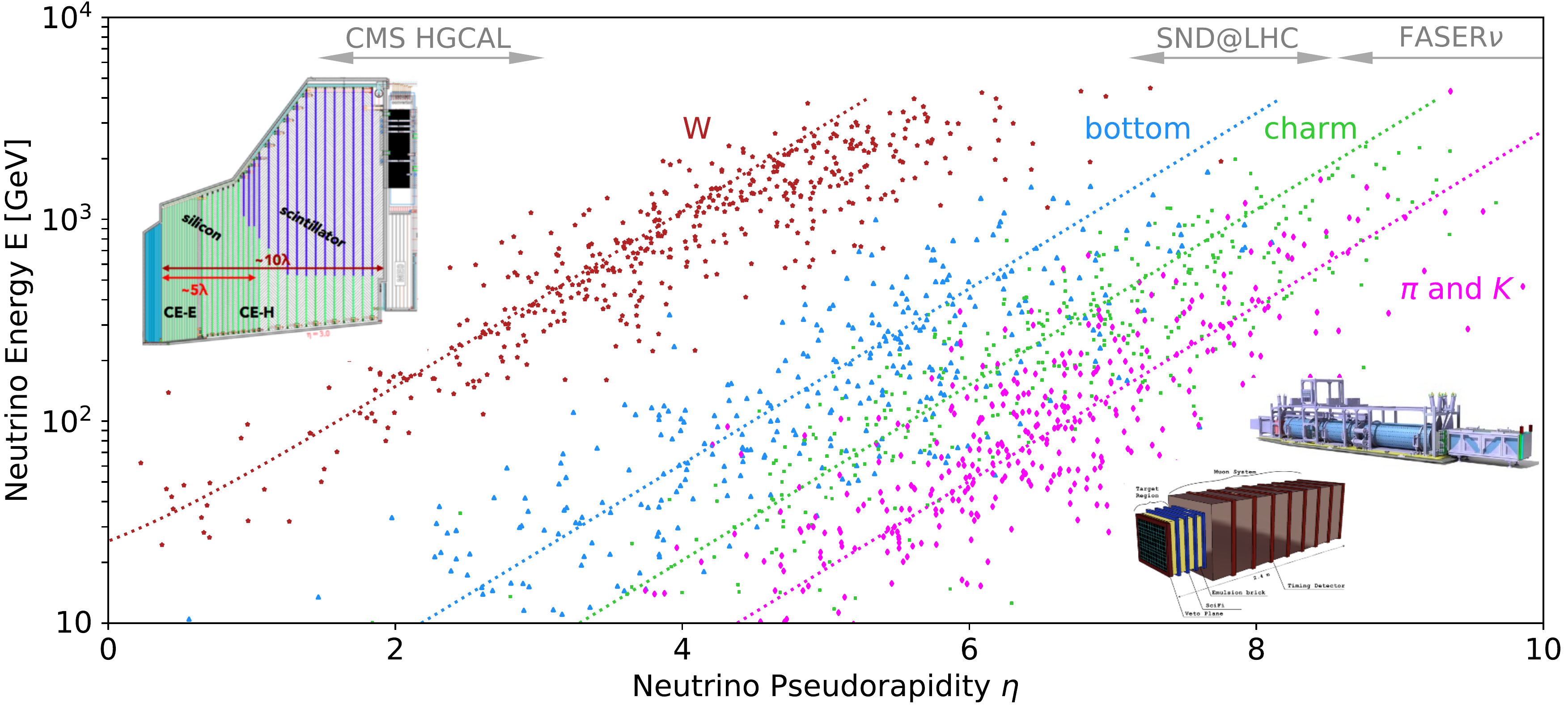}
    \caption{Sensitivities of various LHC based experiments to the LHC neutrino flux. Sources considered are neutrinos from $W$ boson, bottom, charm as well as $\pi$ and $K$ meson decays. Figure taken from~\cite{Foldenauer:2021gkm}.}
    \label{fig:nuflux}
\end{figure}

In~\cref{fig:nuflux} we show the energy distribution of the LHC produced neutrino flux as a function of pseudorapidity $\eta$ for a number of sources. In this figure we illustrate the angular sensitivities of the future forward experiments SND@LHC and FASER$\nu$ alongside the upgraded CMS endcap high-granularity calorimeter (HGCAL).
As can be seen, the forward physics experiments are excellently suited to detect a large fraction of the flux of high-energy neutrinos from meson decays. However, there is still a large unexploited reservoir of neutrinos produced from $W$ decays in the process $W\to \nu_\mu\,\mu$, which is in the angular acceptance of the CMS HGCAL. This serves as a strong motivation to investigate the HGCAL sensitivity to detect these neutrinos via scattering. 

\subsection{How to detect neutrinos at CMS}

Inspired by our study of detecting ULDM scattering via appearing displaced jet signatures at LHC multipurpose experiments, we are investigating appearing neutral jets from neutrino-nucleus scattering in the calorimeter material as a detection signature. Isolating these neutrino scattering events at a hadron collider is extremely challenging due to the large background of neutral hadrons from underlying event and pile-up. However, the CMS HGCAL upgrade has some very advantageous features to tackle this background.

\begin{figure}[t]
  \centering
  \includegraphics[width=0.5\textwidth]{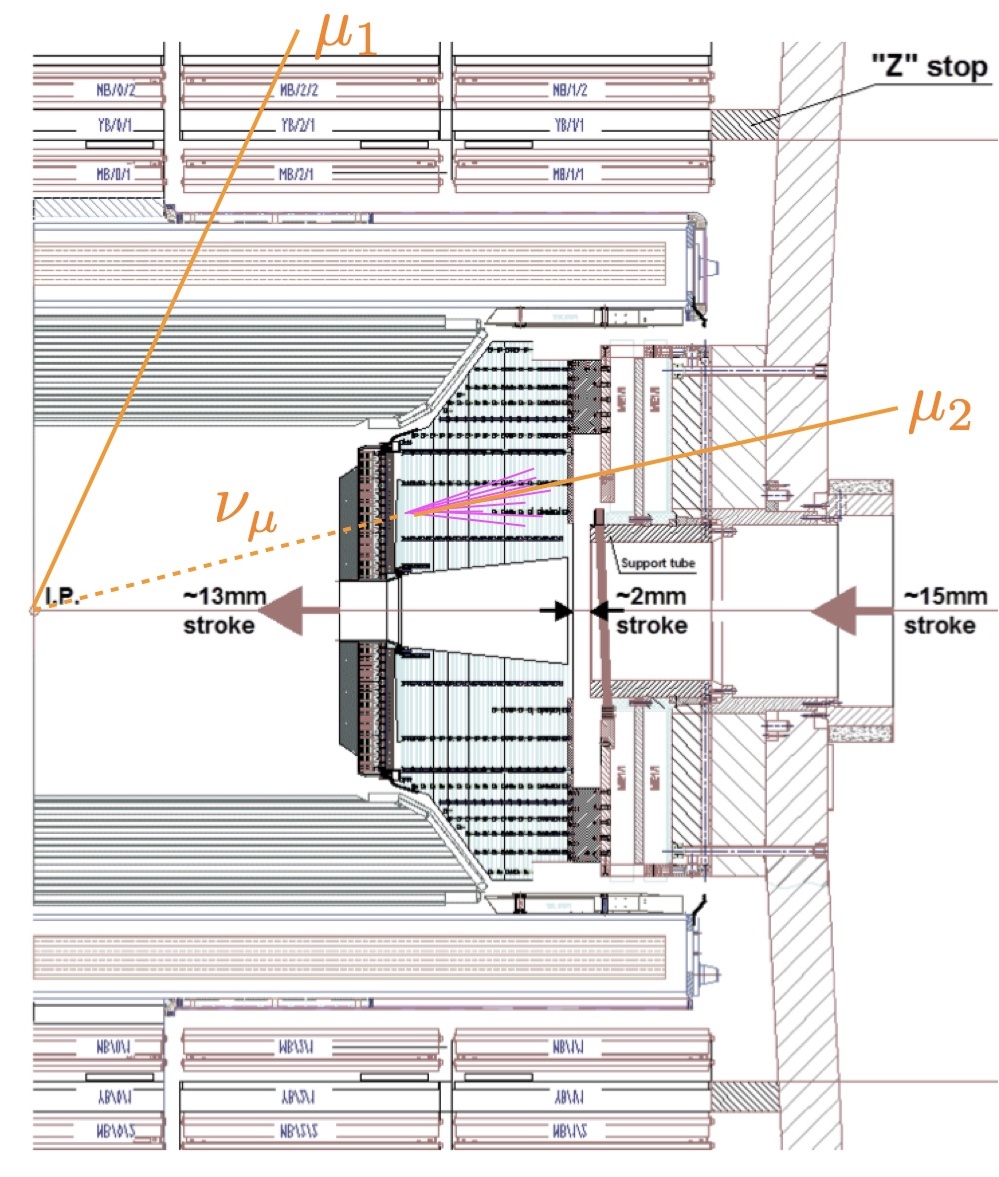}%
  \includegraphics[width=0.5\textwidth]{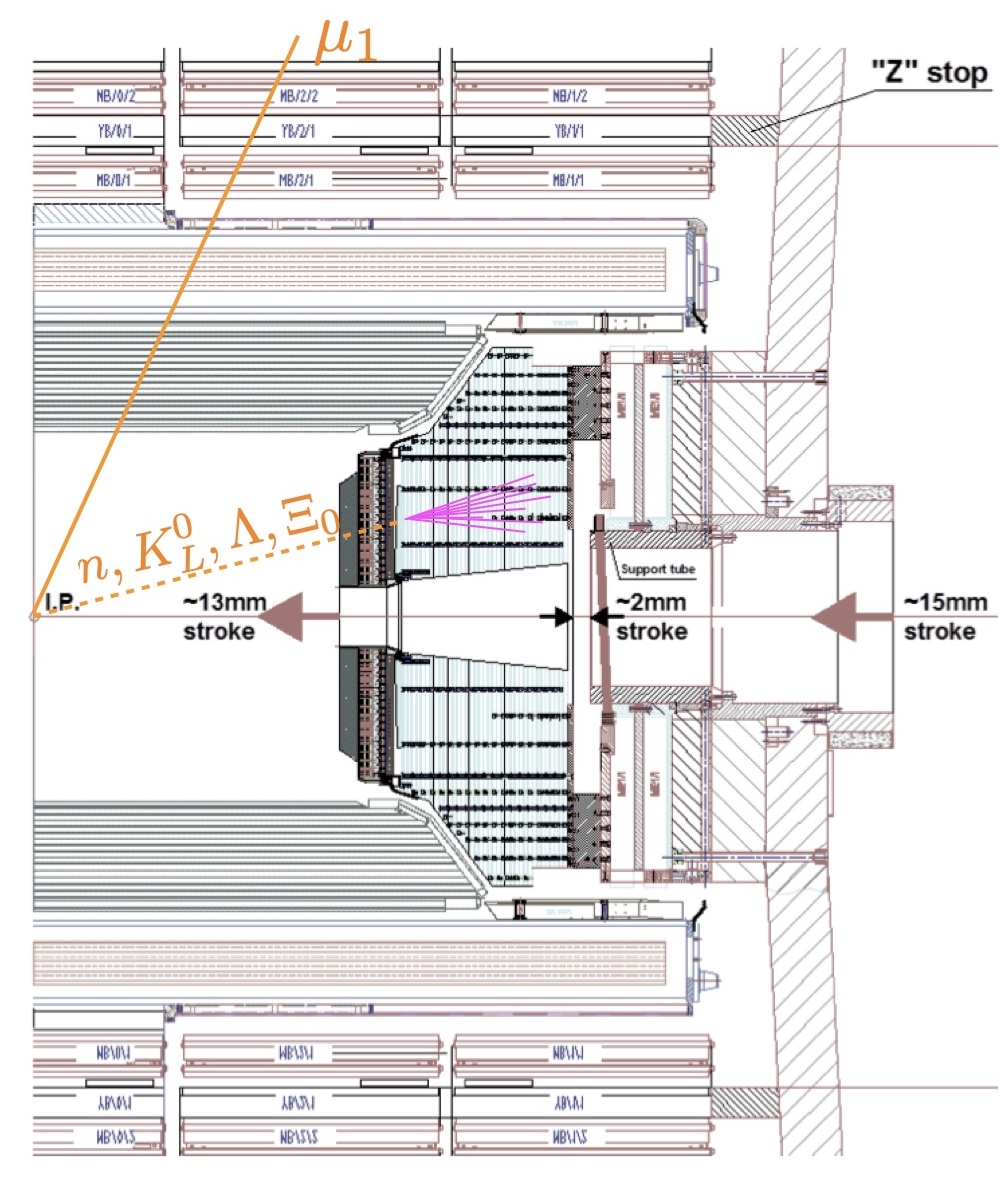}
  \caption{Signal (left) and background (right) topology for neutrino scattering in the CMS detector leading to an appearing jet in the HGCAL.}
  \label{fig:hgcal}
\end{figure}

The HGCAL is a sampling detector with an angular coverage of $1.5<|\eta|<3.0$. It consists mainly of silicon as active material with a size of 0.5 cm$^2$ - 1 cm$^2$ of the individual cells in the transverse plane~\cite{CMS:2017jpq}. Such a finely granulated detector enables a high-resolution measurement of the lateral development of electromagnetic showers. This is crucial for a good two-shower separation  and the observation of narrow jets.
Furthermore, the HGCAL has excellent timing capabilities and is able to operate with a  timing window of 90 ps to remove pile-up, corresponding to a path length difference travelled by relativistic particles of $\Delta_l \sim 2.7$ cm. A further step to remove pile-up even more in the future is a novel proposed collision technique for the HL-LHC, the so-called {\it crab kissing}~\cite{Fartoukh:2014nga}. Here, the bunches are colliding at a crossing angle. In order to achieve maximum spatial overlap, they pass a radio frequency cavity giving the tail and head of the bunches a kick so that they rotate.
With a typical spatial extension of 31.4 cm of the bunch and about $130$ expected pile-up events per bunch crossing~\cite{MedinaMedrano:2301928}, the timing resolution will allow to reduce the number of pile-up events per bunch crossing on average to
\begin{equation}
    N_{\rm pu} \approx \frac{2.7}{31.4} \ 130 \sim 11 \,.
\end{equation}

The neutrino signal which we are focusing on is given by $\nu_\mu$ produced in a $W$ decay in conjunction with a charged muon, and the subsequent scattering of the neutrino with the detector.
The scattering of the $\nu_\mu$ with the detector occurs mostly via the charged current interaction at high energies. There the neutrino converts into a charged lepton and the scattered quark will confine into a jet. Hence, the resulting characteristic signature is a charged lepton together with a single energetic jet produced in a single displaced vertex in the calorimeter with no tracks pointing back to the interaction point.

The signal is hence characterised by the  process,
\begin{align}
    \text{Production:} &\quad 
 qq' \to W\to \mu_1~\nu_\mu\,,  \\
\text{Scattering:} &\quad
\nu_\mu N \to \mu_2 + \text{jet}  \,,
\label{eq:signal}
\end{align}
where a prompt primary muon $\mu_1$ originates from production and a secondary displaced muon $\mu_2$ appears in the scattering. This signal topology in the CMS HGCAL is illustrated in the left panel of~\cref{fig:hgcal}.
To determine the neutrino scattering rate with the detector, we have used the deep-inelastic neutrino interaction cross-sections at leading order as derived in Ref.~\cite{FASER:2019dxq}.

The leading backgrounds to this signal are  isolated muons produced in conjunction with  a long-lived neutral hadron interacting in the calorimeter. 
Potential candidates to mimic the displaced jet of a neutrino signal are neutrons, $K_L^0$, $\Lambda$, or $\Xi^0$. These hadrons are produced copiously in hard scattering, the underlying event or originate from  pile-up. Since these are very likely to interact in the hadronic calorimeter our main background consists of 
\begin{itemize}
    \item[i)] The decay of a $W$-boson into leptons in association with additional neutral hadrons:
    \begin{equation}
        qq' \to W + {\rm QCD}  \to \mu_1 + {\rm QCD} \,,
    \end{equation}
    \item[ii)] Heavy quark production with additional neutral hadrons, with the heavy flavour decaying into muons:
    \begin{equation}
        qq' \to b/c + {\rm QCD}  \to  \mu_1 + {\rm QCD}\,.
    \end{equation}
\end{itemize}
In these processes an appearing jet is produced form the scattering of the neutral hadron in the calorimeter:
\begin{equation}
    \text{neutral hadron} + N \to \text{jet}.
    \label{eq:bkg0}
\end{equation}
The corresponding background topology in the CMS HGCAL is shown in the right panel of~\cref{fig:hgcal}. As can be seen immediately a striking discriminating feature is the secondary muon in the signal.

\begin{figure}[t]
    \centering
    \includegraphics[width=0.5\textwidth]{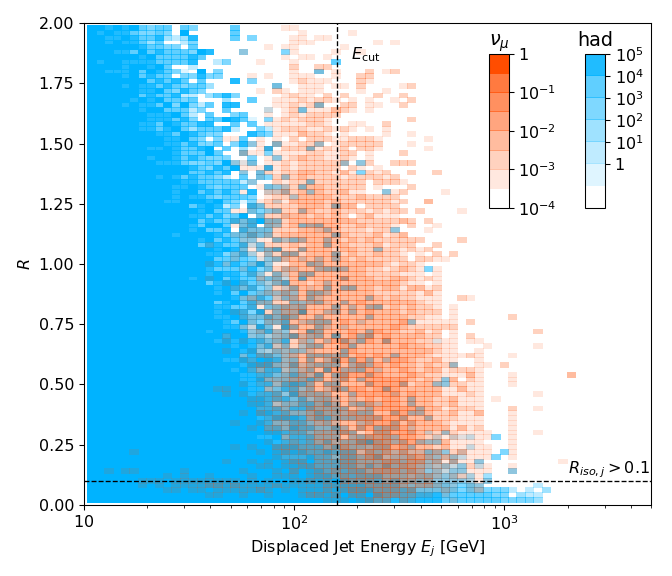}%
    \includegraphics[width=0.5\textwidth]{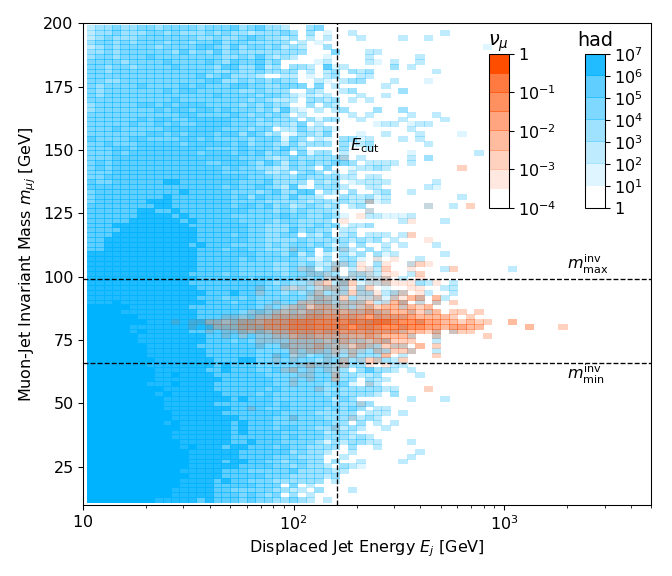}
    \caption{ Distribution of signal (red) and background (blue) events in the jet energy versus $R$ plane (left) and in the jet energy versus muon-jet invariant mass plane (right). The left histogram shows the selection after applying the isolated muon cut and the right one after applying the jet isolation cut.
    Figures taken from~\cite{Foldenauer:2021gkm}.}
    \label{fig:distr}
\end{figure}

In our study we have simulated both signal and background for a data set of 3 ab$^{-1}$ at leading order including parton shower and hadronisation with \textsc{Pythia~8.2}~\cite{Sjostrand:2014zea}. Our analysis strategy is then as follows:

\begin{figure}[t]
{\centering
\begin{minipage}{0.45\textwidth}
  \centering\renewcommand\arraystretch{1.3}
  \begin{tabularx}{\textwidth}{X | X | X }
\hline
\hline
 Cuts & Hadrons & Neutrinos\\
\hline
isolated muon & $1.02\cdot 10^{11}$ & 7.59 \\
isolated jet & $8.63\cdot 10^{10}$ & 7.05 \\
$W$ mass  &  $1.92\cdot 10^9$ & 6.55 \\
secondary muon  & $3.49 \cdot 10^5$  & 5.48 \\ \hline
$E_\mathrm{j}> 160$ GeV &  3.52 & 3.60 \\
\hline
\hline
\end{tabularx}
\end{minipage}%
\begin{minipage}{0.55\textwidth}
  \centering
  \includegraphics[width=\textwidth]{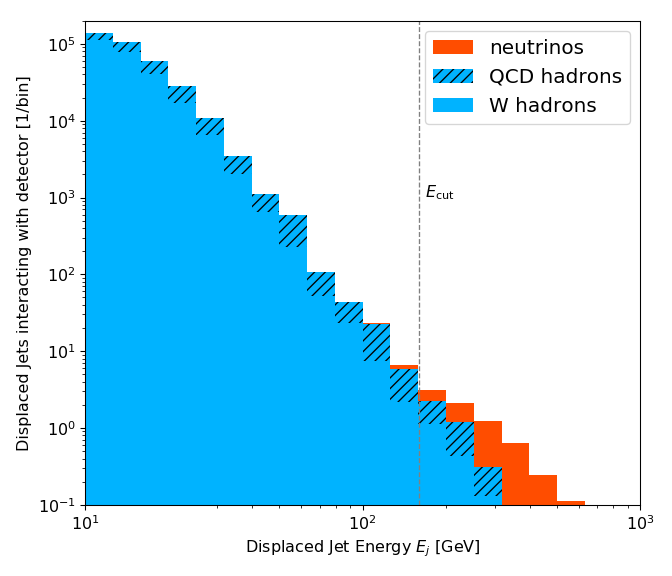}
\end{minipage}
}
\caption{(Left) Analysis cut flow table. (Right) Final histogram of signal (red) and background (red) events after all cuts. Taken from~\cite{Foldenauer:2021gkm}.}
\label{fig:cutflow}
\end{figure}

\begin{itemize}
    \item[\textit{a.}] \textit{Isolated central muon:} We start by requiring a central, isolated primary muon with
    \begin{equation}
       R_{\text{iso},\mu_1}>0.1, \quad p_{T,\mu_1}>20~\text{GeV}, \quad |\eta_{\mu_1}|<2.4~. 
    \end{equation}
    \item[\textit{b.}] \textit{Isolated jet:}
    To reject the fake background jets we use that energetic neutral hadrons are mostly produced as part of an energetic, and hence collimated, hadronic shower. To distinguish between jets from neutral hadrons and neutrinos, we therefore apply a jet isolation cut,
    \begin{equation}
         R_{{\rm iso},j}>0.1\, .
    \end{equation}
    \item[\textit{c.}] \textit{W mass cut:} Since our signal is consisting of a muon and a muon-neutrino originating from a $W$ decay, we require the neutrino jet and the primary muon to reconstruct the $W$ mass. Defining $(p_{\mu_1}+p_\nu)^2=m_{\mu\nu}^2$ we hence require
    \begin{equation}
        66~\text{GeV} < m_{\mu\nu} < 99~\text{GeV} \,.
    \end{equation}
    \item[\textit{d.}] \textit{Secondary muon:}
    We can search for the additional secondary muon as an extra handle to isolate the signal with the CMS endcap muon stations located behind the HGCAL covering $1.2 \!<\! |\eta| \!<\! 2.4$.
    Hence, we require the hardest muon originating from the detector material collision to carry away 
    \begin{equation}
        E_{\mu_2}/E_\mathrm{j}>0.33\,.
        \label{eq:secondary}
    \end{equation}
    \item[\textit{e.}] \textit {Energy cut:}
    Since neutrinos produced in $W$ boson decays are expected to be more energetic than the neutral hadron background, we only consider events with displaced jet energies of
    \begin{equation}
        E_{\rm cut}>160 \text{GeV}\,.
    \end{equation}
\end{itemize}

In~\cref{fig:distr} we show the signal versus background distributions both in the jet energy versus $R$ and muon-jet invariant mass plane. For illustration we also display the jet-isolation, the invariant mass reconstruction and the energy cut. As can be seen these cuts are well suited to separate the signal from the background. However, the most important selection criterion is the detection of the energetic secondary muon, which can suppress the background by almost four orders of magnitude. The full cut flow table is shown on the left of~\cref{fig:cutflow}, which shows that this analysis is able to reduce the initial huge hadron background to a similar level of remaining neutrino signal events.
In the right panel of~\cref{fig:cutflow} we show the final histogram of signal and background events as a function of jet energy. In summary, we can see that in terms of a proof-of-principle strategy this analysis is capable of detecting neutrino scattering at the CMS HGCAL.

\section{Conclusions}
\label{sec:concl}

Summarising our results presented in the last two sections, we can conclude that appearing displaced jets are a promising signature to search for the scattering of light elusive particles in LHC multipurpose experiments.

In particular, in~\cref{sec:uldm} we have argued that high-energy appearing jets in the ATLAS calorimeters are a sensitive probe of large parts of the parameter space of a shift-symmetric light pseudo-scalar particle $a$ coupled to the SM via a new heavy mediator. This is mainly due to the momentum-enhanced scattering cross section at high-energy colliders over low-energy precision probes. 

In~\cref{sec:nus} we presented a proof-of-principle analysis of appearing neutrino-scattering jets at the CMS HGCAL. The crucial handle on the neutrino signal to discriminate it from the large neutral hadronic background is the detection of an energetic secondary muon produced in the charge current neutrino scattering with the detector material.

In the future, there are still many possible improvements to be made to these search strategies. In particular, one can apply machine learning techniques to study the lateral shower development in the detectors to better discriminate BSM and neutrino jets from hadronic background jets. For neutrino scattering, it is still worth investigating the prospects of detecting scattering from neutrinos produced in bottom and charm flavoured meson decays. Furthermore, one could search for neutrino scattering directly in the central muon stations.

\bigskip

\paragraph{Acknowledgements.}
The author would like to express special thanks to the Mainz Institute for Theoretical Physics (MITP) of the Cluster of Excellence PRISMA+ (Project ID 39083149), for its hospitality and support.
The work of PF was partially supported by the UKRI Future Leaders Fellowship DARKMAP and the Spanish Agencia Estatal de Investigaci\'on through the grants PID2021-125331NB-I00 and CEX2020-001007-S, funded by
\\
MCIN/AEI/10.13039/501100011033.


\end{document}